\documentclass[a4paper,11pt]{article}
\pdfoutput=1 

\usepackage{jheparxiv}

\usepackage[T1]{fontenc} 

\usepackage{tensor}
\usepackage{amsmath}
\usepackage{amssymb}
\usepackage{mathrsfs}
\usepackage{mathtools}
\usepackage{bbm}
\usepackage{graphicx}
\usepackage{stmaryrd}
\usepackage{twistor}
\usepackage{subcaption}
\usepackage{xcolor}
\usepackage{empheq}

\newcommand{\eps}{\epsilon}
\newcommand{\veps}{\varepsilon}
\newcommand{\bo}{\mathbf{o}}
\newcommand{\bi}{\boldsymbol{\iota}}

\newcommand{\sT}{\mathsf{T}}
\renewcommand{\sc}{\mathsf{c}}

\newcommand{\sa}{\mathsf{a}}
\newcommand{\sH}{\mathsf{H}}
\newcommand{\ta}{\mathrm{a}}
\renewcommand{\sT}{\mathsf{T}}
\newcommand{\sK}{\mathsf{K}}

\title{Gluon scattering on the self-dual dyon}

\author[a]{Tim Adamo,}
\author[b]{Giuseppe Bogna,}
\author[b]{Lionel Mason,}
\author[c]{Atul Sharma}

\affiliation[a]{School of Mathematics and Maxwell Institute for Mathematical Sciences,\\
	University of Edinburgh, EH9 3FD, UK\vspace{0.1cm}}
\emailAdd{t.adamo@ed.ac.uk}

\affiliation[b]{The Mathematical Institute, University of Oxford,\\ Woodstock Road, Oxford OX2 6GG, UK\vspace{0.1cm}}
\emailAdd{giuseppe.bogna@maths.ox.ac.uk}
\emailAdd{lmason@maths.ox.ac.uk}

\affiliation[c]{Center for the Fundamental Laws of Nature \& Black Hole Initiative,\\Harvard University, Cambridge, MA, 02138, USA \vspace{0.1cm}}
\emailAdd{atulsharma@fas.harvard.edu}

\abstract{The computation of scattering amplitudes in the presence of non-trivial background gauge fields is an important but extremely difficult problem in quantum field theory. In even the simplest backgrounds, obtaining explicit formulae for processes involving more than a few external particles is often intractable. Recently, it has been shown that remarkable progress can be made by considering background fields which are chiral in nature. In this paper, we obtain a compact expression for the tree-level, maximal helicity violating (MHV) scattering amplitude of an arbitrary number of gluons in the background of a self-dual dyon. This is a Cartan-valued, complex gauge field sourced by a point particle with equal electric and magnetic charges, and can be viewed as the self-dual version of a Coulomb field. Twistor theory enables us to manifest the underlying integrability of the self-dual dyon, trivializing the perturbative expansion in the MHV sector. The formula contains a single position-space integral over a spatial slice, which can be evaluated explicitly in simple cases. As an application of the formula, we show that the holomorphic collinear splitting functions of gluons in the self-dual dyon background are un-deformed from a trivial background, meaning that holomorphic celestial OPE coefficients and the associated chiral algebra are similarly un-deformed. We also comment on extensions of our MHV formula to the full tree-level gluon S-matrix.}

\begin{document}
	
	\maketitle
	\flushbottom
	
\section{Introduction}\label{sec:intro}

The computation of scattering amplitudes in the presence of non-trivial background fields -- solutions to the classical equations of motion which are treated non-perturbatively -- is important for the study of physical scenarios where there is a large separation of scales. In the context of abelian and non-abelian gauge theories, such `strong-field' scattering amplitudes play a key role in the study of non-linear regimes of QED due to intense electromagnetic fields (cf., \cite{DiPiazza:2011tq,Lai:2014nma,Fedotov:2022ely}) and in the high-energy, high-density regime of QCD describing heavy ion collisions (cf., \cite{Iancu:2002xk,Iancu:2003xm,Gelis:2010nm}). However, actually computing strong-field amplitudes in even the simplest of background field configurations is an extremely difficult problem. 

The standard approach to this is via the background field formalism (cf., ~\cite{Furry:1951zz,DeWitt:1967ub,tHooft:1975uxh,Abbott:1981ke}), which requires establishing the background-coupled Feynman rules in a form amenable to explicit calculations. Even for highly-symmetric backgrounds where there are closed-form solutions for background-coupled free fields, such as electromagnetic plane waves in QED~\cite{Wolkow:1935zz,Seipt:2017ckc} or gluonic shockwaves in QCD~\cite{Balitsky:1995ub,Caron-Huot:2013fea}, the resulting amplitudes are no longer rational functions (as in a trivial background) and typically involve position space integrals which cannot be evaluated analytically and whose number grows with the number of external particles\footnote{This has led to the development of strong background models which are amenable to numerical evaluation. A prominent example is the locally constant field approximation (LCFA) in strong-field QED~\cite{Reiss:1962nhe,Ritus:1985vta,Baier:1998vh}; while suitable for numerical treatments, there are many physical shortcomings of the LCFA (see~\cite{Harvey:2014qla,DiPiazza:2017raw,Ilderton:2018nws,DiPiazza:2018bfu,Adamo:2021jxz}).}. Consequently, the precision-frontiers for strong-field scattering amplitudes (at least, for analytic computations) are for low numbers of external particles, anywhere from 2- to 4-points depending on the complexity of the background (see~\cite{Fedotov:2022ely} for a recent review containing the precision frontiers for strong-field QED, for instance). This is in sharp contrast to the state of affairs for scattering amplitudes in a trivial background, where many \emph{all-multiplicity} formulae exist, including the famous Parke-Taylor formula for maximal helicity violating gluon scattering in Yang-Mills theory~\cite{Parke:1986gb}.

Obtaining high-multiplicity (in external particle number) strong-field scattering amplitudes is not just a theoretical challenge, though. While adding additional external particles leads to more powers of the small coupling constant indexing perturbations around the background, the increase in multiplicity also leads to more insertions of background-dressed propagators and wavefunctions, each carrying explicit dependence on the background. For sufficiently `strong' background fields, these additional insertions can overwhelm the small coupling constant, leading to high-multiplicity processes dominating low-multiplicity ones (cf., \cite{DiPiazza:2011tq,King:2015tba,Seipt:2017ckc,Fedotov:2022ely}). Consequently, finding analytic approaches to high-multiplicity strong-field scattering has both experimental as well as theoretical imperatives\footnote{It should be noted that some all-multiplicity expressions for scattering have been obtained using the worldline formalism in strong-field QED (e.g., multi-photon production from electron scattering in a plane wave)~\cite{Edwards:2021vhg,Schubert:2023gsl,Ahmadiniaz:2023jwd,Copinger:2023ctz,Copinger:2024twl}. However, these are written as `master formulae' from which it is still rather complicated to extract individual S-matrix elements.}. 

\medskip

To approach the daunting task of computing high-multiplicity strong-field scattering amplitudes, one can first look for background fields which are simpler than those of immediate experimental interest but which nevertheless capture many of the essential features of strong-field scattering. A particularly interesting class of such examples is provided by \emph{self-dual} background field configurations in gauge theory and gravity. These are solutions of the classical equations of motion whose gauge-covariant field strength is self-dual when viewed as a 2-form in four-dimensions. While intrinsically chiral, and thus complex in Lorentzian signature, such self-dual backgrounds often serve as simplified toy-models of more realistic strong-fields (e.g., plane waves, Coulomb charges or black holes) and scattering in these backgrounds still exhibits many of the complicated structures associated with generic strong-field amplitudes, such as a lack of rational functions, tails and memory effects. 

The self-dual (SD) sectors are classically integrable~\cite{Ward:1977ta,Penrose:1976js,Mason:1991rf}, enabling powerful methods based on twistor theory to be brought to bear on the computation of scattering amplitudes SD backgrounds. Twistor methods have now been used to derive \emph{all-multiplicity} formulae for tree-level, maximal helicity violating (MHV) gluon and graviton scattering amplitudes on any SD radiative background in Yang-Mills theory and general relativity~\cite{Adamo:2020syc,Adamo:2020yzi,Adamo:2022mev}. These backgrounds are determined by their characteristic data at null infinity, a freely-specified (spin- and conformally-weighted) complex function of three variables. In addition, twistor theory makes it possible to conjecture formulae for the \emph{full} tree-level S-matrices of gluons and gravitons on such backgrounds; while these formulae have not yet been proven beyond the MHV sector, they pass several non-trivial consistency checks. It is even possible to go beyond tree-level and fully on-shell quantities in such backgrounds, with one-loop all-positive-helicity amplitudes and tree-level form factors in Yang-Mills having also been computed to arbitrary multiplicity in SD radiative gauge field backgrounds~\cite{Bogna:2023bbd}.

\medskip

In this paper, we turn to the computation of gluon scattering amplitudes on another SD gauge field background: the \emph{self-dual dyon} (SDD). This is a solution of the Maxwell equations composed of a Coulomb charge and magnetic monopole, with the electric and magnetic charges tuned so that the resulting field strength is purely self-dual. On Minkowski spacetime $\R^{1,3}$, the SDD is a complex-valued gauge field, but on Euclidean $\R^4$ or split signature $\R^{2,2}$ it is real-valued. Furthermore, it can be trivially lifted to a SD solution of the Yang-Mills equations by embedding in a Cartan subalgebra of the non-abelian gauge group. 

On the one hand, the SDD might appear to be a simpler SD background than any SD radiative gauge field, since it has no functional degrees of freedom. On the other hand, it is a stationary solution and essentially non-radiative: it can be viewed as the gauge field sourced by a Cartan-valued dyonic point-particle, moving on a time-like worldline in flat space~\cite{Huang:2019cja,Emond:2020lwi}. Thus, scattering on the SDD background is the SD prototype of Coulomb scattering. The standard Coulomb problem is rife with complications\footnote{Indeed, even the perturbative description of Coulomb scattering requires modified asymptotics for the wavefunctions~\cite{Lippstreu:2023vvg}.}: the relativistic Coulomb wave equation can only be solved using separation of variables and a partial wave expansion, with radial modes given by confluent hypergeometric functions. Even the 2-point amplitude (which is only sensitive to the asymptotics of the Coulomb wavefunction) is given by an infinite sum of terms controlled by the radial action~\cite{Kol:2021jjc,Adamo:2023cfp}. A similar story is true for scattering on a generic (non-self-dual) dyon background~\cite{Schwinger:1976fr,Boulware:1976tv,Shnir:2005vvi,Kol:2021jjc}.

However, for the SDD a miraculous simplification occurs already at the level of the massless background-coupled wavefunctions. In~\cite{Adamo:2023fbj}, we showed that in this case charged Killing spinors exist which allow us to write gluon wavefunctions in the SDD background in a closed compact form (i.e., without a partial wave expansion) as \emph{quasi-momentum eigenstates}. These have the property of being everywhere regular on the celestial sphere and at the origin. Furthermore, the charge of the gluons with respect to the Cartan-valued background is quantized due to the topological non-triviality of the dyon, but when this charge is set to zero the quasi-momentum eigenstates smoothly reduce to the usual momentum eigenstates on a trivial background. Armed with these simple external states, the tree-level 2-point gluon amplitude on a SDD background was calculated explicitly in~\cite{Adamo:2023fbj}, resulting in a surprisingly simple closed-form expression.

In this paper, we compute \emph{all-multiplicity} gluon scattering amplitudes in the background of a SDD, focusing on the MHV helicity sector. Exploiting the twistor description of the SDD~\cite{Sparling:1979,Hughston:1979tq}, gluon quasi-momentum eigenstates and the generating functional for MHV amplitudes in the background can be lifted from spacetime to twistor space, where its perturbative expansion is essentially straightforward. This allows us to obtain a formula for the MHV amplitude in terms of an integral over a single spatial point. For instance, when all of the scattered gluons are `minimal' (in the sense that they have minimal growth at large distances), the $n$-point, colour-ordered MHV amplitude (where gluons $r,s$ have negative helicity and all others have positive helicity) is 
\begin{multline}\label{MHVpreview}
\cA^{\mathrm{MHV}}_{n}=2\pi\,\rg^{n-2}\,\delta\!\left(\sum_{i=1}^{n}\omega_i\right) \delta\!\left(\sum_{i=1}^ne_i\right)\,\frac{\la r\,s\ra^4}{\la1\,2\ra\,\la2\,3\ra\cdots\la n\,1\ra} \\
\times \int \d^{3}\vec{x}\,\e^{\im\,\vec{k}\cdot\vec{x}} \prod_{a\in\mathfrak{n}_+}\left(\frac{r}{1+|\zeta|^2}\right)^{e_a}\left(\bar{\zeta}\,z_a+1\right)^{2e_a}\prod_{b\in\mathfrak{n}_-}\left(\frac{r}{1+|\zeta|^2}\right)^{-e_b}\left(\zeta-z_b\right)^{-2e_b}\,,
\end{multline}
where $\rg$ is the Yang-Mills coupling constant, $\mathfrak{n}_{\pm}$ are the sets of positive/negative charge gluons (with respect to the background) and $e_i$, $\omega_i$ are the charge and frequency of gluon $i$, whose (complex) on-shell 4-momentum is parametrized as
\be\label{kos}
k_{i}^{\alpha\dot\alpha}=\kappa^{\alpha}_i\,\tilde{\kappa}_i^{\dot\alpha}\,, \qquad \kappa_i^{\alpha}=(1,z_i)\,, \quad \tilde{\kappa}_i^{\dot\alpha}=\frac{\sqrt{2}\,\omega_i}{1+|z_i|^2}\,(1,\bar{z}_i)\,.
\ee
The rational factor in the first line is constructed from the usual invariants $\la i\,j\ra\equiv \kappa_i^{\alpha}\kappa_{j\,\alpha}$, and the integral in the second line is over $(0,\infty)\times S^2$, where $r\in(0,\infty)$ and $(\zeta,\bar{\zeta})$ are stereographic coordinates on the 2-sphere, and $\vec{k}:=\sum_{i=1}^{n}\vec{k}_i$. 

These integrals can be performed explicitly in many cases (as we demonstrate), but even the un-integrated form is remarkable: background field theory suggests that for a QFT with cubic interactions, such as Yang-Mills, in a stationary background, one should expect a $n$-point tree-amplitude should have $3(n-2)$ position-space Feynman integrals. 

Armed with explicit formulae such as \eqref{MHVpreview}, it becomes possible to explore the infrared (IR) structures of gluon scattering on the SDD, particularly holomorphic collinear limits. These are equivalent, after performing a Mellin transform in the frequencies of the external gluons, to celestial OPE coefficients~\cite{Fan:2019emx,Pate:2019lpp,Himwich:2021dau}, giving important low-energy data which can be recast in the language of a two-dimensional conformal field theory (CFT). Upon performing a conformally-soft mode expansion, these celestial OPE coefficients encode a chiral algebra associated to positive helicity gluons living in the SDD background. We find that the holomorphic collinear limits -- or equivalently, the holomorphic celestial OPEs -- and associated chiral algebra on the SDD are equivalent to those on a trivial background. This is due to the essentially abelian nature of the background, and is not in tension with other observations on holomorphic chiral algebras on self-dual backgrounds (e.g., \cite{Garner:2023izn,Costello:2023hmi,Adamo:2023zeh}). 

	\paragraph{Summary of the paper:} In Section \ref{sec:twistor}, we introduce the necessary tools to describe the self-dual dyon in the language of twistor theory. We end the Section with a discussion of the scattering states: these are plane-wave solutions to the linearised equations of motion around the self-dual dyon background of the appropriate spin and are thus dubbed \emph{quasi-momentum eigenstates}, even though the underlying background possesses no translational invariance. We discuss the generating functional for the MHV amplitude in Section \ref{sec:mhv} and derive an integral formula for the scattering amplitude of $n$ quasi-momentum eigenstates around the self-dual dyon background. We also include some examples at low points and for special charge configurations where the integral can be performed explicitly. In Section \ref{sec:ope}, we use the general formula to obtain the celestial OPE in the presence of a self-dual dyon and find that the non-trivial background does not deform the chiral symmetry algebra that arises around the vacuum. We give some concluding remarks and address possible open questions in Section \ref{sec:concl}. Appendix \ref{app:integration} contains details about integration of the MHV amplitude in special charge configurations.

	
\section{Twistor theory for the self-dual dyon}\label{sec:twistor}

The self-dual dyon (SDD) is a solution of Maxwell's equations in four-dimensions with a purely self-dual field strength. This means that it admits a description in terms of twistor theory, and the relative simplicity of this description will underpin our later calculation of all multiplicity gluon scattering amplitudes in the SDD background. In this section, we review the SDD field, describe its twistor theory and investigate how it encodes solutions of the charged, background-coupled zero-rest-mass equations.

\medskip 
	
We begin by briefly fixing our notation conventions (broadly following~\cite{Adamo:2017qyl}). Let $x^a$, $a=0,\ldots,3$ be Cartesian coordinates on (complexified) Minkowski space $\M$; these are encoded in the 2-spinor variables
	\be
	x^{\alpha\dot\alpha}=\frac{1}{\sqrt{2}}\left(\begin{array}{cc}
		x^0+ x^3& x^1-\im x^2  \\
		x^1+\im x^2& x^0-x^3 
	\end{array}\right)\,.
	\ee
Two-spinor indices are raised and lowered using the the two-dimensional Levi-Civita symbols $\epsilon_{\alpha\beta}$, $\epsilon_{\dot\alpha\dot\beta}$, etc.. We employ the usual notation for 2-spinor SL$(2,\C)$-invariants: $\la a\,b\ra:=a^{\alpha}b_{\alpha}=\epsilon^{\alpha\beta}a_{\beta}b_{\alpha}$, and $[a\,b]:=a^{\dot\alpha}b_{\dot\alpha}=\epsilon_{\dot\alpha\dot\beta}a_{\dot\beta}b_{\dot\alpha}$.

In these variables, let $T^{\alpha\dot\alpha}\coloneqq\text{diag}(1,1)/\sqrt{2}$ denote the unit, future-pointing timelike vector. As usual, this vector can be used to project the coordinates onto the spatial 3-slices by defining
	\be
x^{\alpha\beta}\coloneqq\,\epsilon_{\dot\alpha\dot\beta}\,T^{(\alpha|\dot\alpha|}\,x^{\beta)\dot\beta}=-\frac{1}{2}\left(\begin{array}{cc}
		-x^1+\im x^2&x^3  \\
		x^3&x^1+\im x^2 
	\end{array}\right)\,,\label{eq:3d-x_from_4d-x}
	\ee
so that $x^{\alpha\beta}$ are coordinates on $\R^3$, whilst $x^0=T^{\alpha\dot\alpha} x_{\alpha\dot\alpha}$ is the coordinate in the direction of $T^{\alpha\dot\alpha}$. The inverse relation
	\be
x^{\alpha\dot\alpha}=x^0\,T^{\alpha\dot\alpha}-2x^{\alpha\beta}\,T_\beta{}^{\dot\alpha}\,,\label{eq:4d-x_from_3d-x}
	\ee
enables us to pass between the $x^{\alpha\dot\alpha}$ and $(x^0,x^{\alpha\beta})$ coordinates.
 
Let $(r,\theta,\phi)$ be standard spherical polar coordinates on $\R^3-0=\R_+\times S^2$. We define the stereographic coordinates $\zeta=\e^{\im\phi}\tan\frac{\theta}{2}$ and $\bar\zeta=\e^{-\im\phi}\tan\frac{\theta}{2}$ on $S^2$, so that $x^{\alpha\beta}$ decomposes as 
	\be
x^{\alpha\beta}=r\,\bi^{(\alpha}\bo^{\beta)}\,,\qquad\bi^{\alpha}\coloneqq\frac{(1,\zeta)}{\sqrt{1+\zeta\bar\zeta}}\,,\qquad \bo^{\alpha}\coloneqq\frac{(\bar\zeta,-1)}{\sqrt{1+\zeta\bar\zeta}}\,.
	\ee
It will also be useful to fix a constant dyad on the bundle of undotted spinors. We take this dyad to be $\iota^\alpha\coloneqq(1,0)$ and $o^\alpha\coloneqq(0,-1)$, normalized so that $\la \iota \,o\ra=1$.

	
\subsection{The self-dual dyon}
	
The self-dual dyon (SDD) in Maxwell theory can be viewed as the gauge field sourced by a point particle carrying electric and magnetic charges of equal magnitude but related by an overall factor of the imaginary unit. This ensures the field strength is self-dual as a 2-form on $\M$. As such, the SDD is a complex gauge field in Lorentzian-real Minkowski space, although it can also be viewed as a real gauge field by taking different signature (e.g., Euclidean or Kleinian) real slices of $\M$. The SDD can be trivially viewed as a solution of non-abelian Yang-Mills theory by embedding it into a Cartan subalgebra of the gauge group. The condition that it is embedded in a Cartan subalgebra ensures that the self-duality condition of the abelian case is preserved in the full non-abelian theory. 

More precisely, let $\sc$ be an element of the Cartan subalgebra of the gauge group (typically, we imagine working with SU$(N)$, but in practice any choice is fine). Setting the magnitude of the electric and magnetic charges in the abelian theory equal to one, the SDD gauge field can be written as
	\begin{equation}
		A\coloneqq\sc\left(\frac{\d x^0}{r}+\ta\right)\,,\qquad \ta\coloneqq\im\,\frac{\zeta\,\d\bar\zeta-\bar\zeta\,\d\zeta}{1+\zeta\bar\zeta}\,.\label{eq:sdd}
	\end{equation}
The first term in $A$ is the Coulombic part, corresponding to the electric field of unit charge in the abelian theory, while $\ta$ gives rise to the magnetic part of the electromagnetic field and is sourced by a unit magnetic monopole. As $A$ points along the Cartan direction $\sc$, the non-Abelian field strength is simply given by
	\be
	F=\d A=\sc\left(\frac{\d x^0\wedge\d r}{r^2}+2\im\,\frac{\d\zeta\wedge\d\bar\zeta}{(1+\zeta\bar\zeta)^2}\right)\,,
	\ee
and thus still satisfies the self-duality equation. In particular, in 2-spinor variables the field strength is
	\begin{equation}
		F_{\alpha\dot\alpha\beta\dot\beta}=\sc\frac{2\im\, x_{\dot\alpha\dot\beta}\epsilon_{\alpha\beta}}{r^3}:=\tilde{F}_{\dot\alpha\dot\beta}\,\epsilon_{\alpha\beta}\,,
	\end{equation}
where $x^{\dot\alpha\dot\beta}\coloneqq-\im\,\epsilon_{\alpha\beta}T^{\alpha(\dot\alpha}x^{|\beta|\dot\beta)}$. Self-duality is manifest in the skew-symmetry of the undotted indices and the fact that $\tilde{F}_{\dot\alpha\dot\beta}=\tilde{F}_{(\dot\alpha\dot\beta)}$.

Other gauges for the SDD gauge potential can be found which manifest other useful features of the field. For instance, in the \emph{Kerr-Schild gauge} the gauge potential is
    \be
    A^\text{KS}=\sc\,\phi\,q\,,\qquad \phi\coloneqq\frac{1}{r}\,,\qquad q\coloneqq\d x^0+\im \d r-2\im r\frac{\bar\zeta}{1+\zeta\bar\zeta}\d\zeta\,.\label{eq:kerr-schild_gauge}
    \ee
Here $\phi$ is a harmonic function, whilst $q=q_a\,\d x^a$ is a 1-form satisfying
    \be
    q^a\,q_a=0\,,\qquad q^a\,\p_aq_b=0\,,
    \ee
so that the dual vector field $q^{a}$ is the tangent vector of a null geodesic in $\M$. Note that this implies that the Kerr-Schild gauge is also a light-cone gauge. This choice of gauge can be used to relate the SDD to the self-dual Taub-NUT metric by the classical double copy~\cite{Monteiro:2014cda,Luna:2015paa}. 

Another natural light-cone gauge can be defined in terms of charged Killing spinors of the SDD field. In particular, the SDD has two charged Killing spinors $\chi_{\pm}^{\alpha}$ of charges $\pm\frac{1}{2}$~\cite{Adamo:2023fbj}:
\be\label{Kspinors}
\chi_+^\alpha=\sqrt{r}\,\bo^\alpha\,,\qquad \chi_-^\alpha=\sqrt{r}\,\bi^\alpha\,,
\ee
obeying
\be\label{KSE}
\partial^{(\alpha}{}_{\dot\alpha}\chi^{\beta)}_{\pm}\pm\frac{\im}{2}\,A^{(\alpha}{}_{\dot\alpha}\,\chi^{\beta)}_{\pm}=0\,,
\ee
for $A_{\alpha\dot\alpha}$ the SDD gauge potential in the `standard' gauge \eqref{eq:sdd}. One can then re-express the SDD in a light-cone gauge adapted to these Killing spinors 
\be
    A^\text{LC}_{\alpha\dot\alpha}=\frac{\iota_\alpha T^\gamma{}_{\dot\alpha}}{r}\left(\frac{\chi_{+\,\gamma}}{\la\iota\,\chi_+\ra}+\frac{\chi_{-\,\gamma}}{\la\iota\,\chi_-\ra}\right)\,,\label{eq:light-cone_gauge}
\ee
where $\iota^{\alpha}=(1,0)$ is the constant spinor from our dyad which the null vector $n^{\alpha\dot\alpha}=\iota^{\alpha}\bar{\iota}^{\dot\alpha}$ for which $n\cdot A^{\mathrm{LC}}=0$. Note that both the Kerr-Schild and light-cone gauges also happen to be Lorenz gauges.

\medskip
	
\paragraph{Exterior derivatives:} It will be useful to know the explicit expressions for the exterior derivatives of the spherical coordinates and Killing spinors of the SDD in terms of Cartesian coordinates. A direct computation shows
	\begin{align}
		\d r&=-2\,\bo_{(\beta}\bi_{\gamma)}\,T^\gamma{}_{\dot\beta}\,\d x^{\beta\dot\beta}\,,\label{eq:dr}\\
		\d\zeta&=-\frac{(1+\zeta\bar\zeta)}{r}\,\bi_\beta\bi_\gamma\, T^\gamma{}_{\dot\beta}\,\d x^{\beta\dot\beta}\,,\label{eq:dzeta}\\
		\d\bar\zeta&=\frac{(1+\zeta\bar\zeta)}{r}\,\bo_\beta\bo_\gamma\, T^\gamma{}_{\dot\beta}\,\d x^{\beta\dot\beta}\label{eq:dbarzeta}\,.
	\end{align}
	These derivatives can be used to compute
	\begin{align}
		\d \bo^\alpha&=-\frac{\im}{2}\,\ta\,\bo^\alpha+\frac{1}{r}\,\bo_\beta\bo_\gamma\, T^\gamma{}_{\dot\beta}\,\d x^{\beta\dot\beta}\,\bi^\alpha\,,\label{eq:dbo}\\
		\d\bi^\alpha&=\frac{\im}{2}\,\ta\,\bi^\alpha+\frac{1}{r}\,\bi_\beta\bi_\gamma\, T^\gamma{}_{\dot\beta}\,\d x^{\beta\dot\beta}\,\bo^\alpha\,,\label{eq:dbi}
	\end{align}
	as well as
	\begin{align}
		\d\chi_+^\alpha&=-\left(\frac{\chi_{+(\beta)}\chi_{-\gamma)}}{r^2}T^\gamma{}_{\dot\beta}\d x^{\beta\dot\beta}+\frac{\im}{2}\ta\right)\chi_+^\alpha+\frac{1}{r^2}\chi_{+\beta}\chi_{+\gamma}T^\gamma{}_{\dot\beta}\d x^{\beta\dot\beta}\chi_-^\alpha\,,\label{eq:dchi+}\\
		\d\chi_-^\alpha&=-\left(\frac{\chi_{+(\beta)}\chi_{-\gamma)}}{r^2}T^\gamma{}_{\dot\beta}\d x^{\beta\dot\beta}-\frac{\im}{2}\ta\right)\chi_-^\alpha+\frac{1}{r^2}\chi_{-\beta}\chi_{-\gamma}T^\gamma{}_{\dot\beta}\d x^{\beta\dot\beta}\chi_+^\alpha\label{eq:dchi-}\,,
	\end{align}
	where $\ta$ is again the magnetic monopole field defined in \eqref{eq:sdd}.

	
\subsection{The twistor quadrille}

We can now describe the SDD in terms of certain holomorphic structures on twistor space. The original description of the SDD in twistor theory was based on a \v{C}ech formulation~\cite{Sparling:1979,Hughston:1979tq}; here we develop a new Dolbeault description which is more suited to our later calculations. The two descriptions are, of course, equivalent thanks to the \v{C}ech-Dolbeault isomorphism. Our focus will be on Euclidean signature to make the discussion more explicit, but it is straightforward to adapt the construction to other signatures or to the complexified setting.

Letting\footnote{We follow the conventions of \cite{Adamo:2017qyl}.} $Z^A=(\mu^{\dot\alpha},\lambda_\alpha)$ be holomorphic, homogeneous coordinates on $\P^3$, 3-dimensional complex projective space. Twistor space is the total space of the fibration $\mathcal{O}(1)\oplus\mathcal{O}(1)\to\P^1$, with $\lambda_\alpha$ being the homogeneous coordinates on the $\P^1$ base:
	\be
	\PT=\{[Z^A]\in\P^3\, |\,\lambda_\alpha\neq0\}\,.
	\ee
Here and in the following, $\mathcal{O}(n)\to\P^1$ is the line bundle whose sections are locally holomorphic functions of homogeneity $n$ in the homogeneous coordinates. We can pull back the line bundles $\mathcal{O}(n)\to\P^1$ by the projection $\PT\to\P^1$, and the resulting line bundles will still be denoted as $\mathcal{O}(n)\to\PT$. The distinction should always be clear from the context.
	
The relation between twistor space and spacetime is encoded in the incidence relations
	\be\label{incidence}
	\mu^{\dot\alpha}=x^{\alpha\dot\alpha}\,\lambda_\alpha\,.
	\ee
For fixed $x\in\M$, these incidence relations describe a twistor line $X\subseteq\PT$; that is, a holomorphically and linearly embedded copy of $\P^1$ inside $\PT$. Notice that we can think of twistor space as $\PT\cong\P^3-\P^1$: the excision of the $\P^1$ subvariety where $\lambda_\alpha=0$ corresponds to the excision of the twistor line associated to the point at spatial infinity. Conversely, for a given $(\mu^{\dot\alpha},\lambda_\alpha)$, the incidence relations describe a totally null self-dual 2-plane in $\M$. That is, two twistor lines $X, Y\subset\PT$ intersect if and only if the corresponding spacetime points $x,y\in\M$ are null-separated. 
	
In Euclidean signature there are no null-separated points, so twistor lines foliate twistor space and Euclidean twistor space is equivalent, as a \emph{real} manifold, to the projective, undotted spinor bundle~\cite{Woodhouse:1985id}: $\PT\cong\R^4\times\P^1$. We can describe the complex structure of $\PT$ in $(x^{\alpha\dot\alpha},\lambda_\alpha)$ coordinates as follows: introduce the quaternionic complex conjugation
	\be
	\begin{split}
 \mu^{\dot\alpha}=(\mu^{\dot 0},\mu^{\dot 1}) \quad & \rightarrow \quad  \hat\mu^{\dot\alpha}\coloneqq(-\bar\mu^{\dot 1},\bar\mu^{\dot 0})\,, \\
 \lambda_\alpha=(\lambda_0,\lambda_1) \quad & \rightarrow \quad \hat\lambda_\alpha\coloneqq(-\bar\lambda_1,\bar\lambda_0)\,.
   \end{split}
	\ee 
This conjugation is quaternionic as it squares to $-1$ and preserves the chirality of the spinors. It can be used to invert the incidence relations 
	\begin{equation}
		x^{\alpha\dot\alpha}=\frac{\hat\mu^{\dot\alpha}\,\lambda^\alpha-\mu^{\dot\alpha}\,\hat\lambda^\alpha}{\langle\lambda\,\hat\lambda\rangle}\,.
	\end{equation}
	A basis of $(0,1)$-forms is then given by
	\be
	\bar e^{\dot\alpha}\coloneqq\frac{\hat \lambda_\alpha \,\d x^{\alpha\dot\alpha}}{\langle \lambda\,\hat\lambda\rangle}\,,\qquad\bar e^0\coloneqq\frac{\D\hat\lambda}{\langle \lambda\,\hat\lambda\rangle^2}\,,
	\ee
	where $\D\hat\lambda\coloneqq\langle\hat\lambda\,\d\hat\lambda\rangle$. The corresponding dual basis of $(0,1)$-vector fields is given by
	\begin{equation}
		\bar\p_{\dot\alpha}\coloneqq\lambda^\alpha\frac{\p}{\p x^{\alpha\dot\alpha}}\,,\qquad \bar\p_0\coloneqq-\langle\lambda\,\hat\lambda\rangle\lambda_\alpha\frac{\p}{\p\hat\lambda_\alpha}\,.
	\end{equation}
	Notice that all forms and vector fields have homogeneity 0 in $\hat\lambda_\alpha$, whilst they have non-trivial weights in $\lambda_\alpha$. Further, the complex structure is adapted to the incidence relations, in the sense that $\bar e^0$ is a $(0,1)$-form on the line $X$, whilst the pull-back of $\bar e^{\dot\alpha}$ to $X$ vanishes.

\medskip

The Ward correspondence~\cite{Ward:1977ta,Newman:1978ze} gives a one-to-one correspondence between self-dual Yang-Mills fields on $\R^4$ with rank-$N$ gauge group and rank-$N$ holomorphic vector bundles $E\to\PT$ such that the restriction to any twistor line, $E|_X$, is topologically trivial\footnote{The particular choice of rank-$N$ gauge group induces various additional conditions on $E$. As the SDD is valued in a Cartan subalgebra of the gauge group, these conditions are not terribly important for our purposes.}. The complex structure $E$ can be encoded locally in a partial connection $\bar{D}:\Omega^0(\PT,\E)\to\Omega^{0,1}(\PT,E)$, and holomorphicity of the vector bundle is translated into the integrability of this partial connection $\bar{D}^2=0$.


It can be shown that in perturbation theory the restriction $\left.E\right|_X$ to any twistor line is not only \emph{topologically} trivial, but also \emph{holomorphically} trivial~\cite{Sparling:1990,Mason:2010yk}. This means that there exists a holomorphic frame $\sH=\sH(x,\lambda):E_X\to\C^{N}$ of homogeneity zero in $\lambda_\alpha$ such that
	\be\label{HolFDef}
	\left.\bar D\right|_X\sH=0\,.
	\ee
The self-dual gauge field on $\R^4$ can be entirely reconstructed from this holomorphic frame: the function $\sH^{-1}\lambda^\alpha\p_{\alpha\dot\alpha}\sH$ has homogeneity $+1$ in $\lambda_\alpha$ and is easily seen to be holomorphic on the line $X$, so that a variation of Liouville's theorem implies that
	\be
	\sH^{-1}\lambda^\alpha\p_{\alpha\dot\alpha}\sH=
	-\im\lambda^\alpha\, A_{\alpha\dot\alpha}(x)\,.\label{eq:ward-gauge-field}
	\ee
The holomorphic frame accounts for spacetime gauge transformations as well, since $\sH$ is defined only up to $\sH(x,\lambda)\to\sH(x,\lambda) g(x)$, $g$ being any smooth function on $\R^4$ valued in the gauge group. This ambiguity is precisely a gauge transformation on the gauge field in \eqref{eq:ward-gauge-field}.
	
For the SDD, $E\to\PT$ is a line bundle known as the \emph{twistor quadrille}~\cite{Sparling:1979,Hughston:1979tq}, which can trivially be uplifted to a Cartan subalgebra of a non-abelian gauge group. Write the partial connection on $E$ locally as $\bar{D}=\dbar+\sa$, for $\sa\in\Omega^{0,1}(\PT,\mathcal{O}\otimes\mathrm{End}E)$. The holomorphic trivialization of $E|_X$ provides an isomorphism between $\mathrm{End}E$ and the Cartan subalgebra, so holomorphicity of $E$ is locally equivalent to $\dbar\sa=0$. 

To recover the SDD on spacetime, take the partial connection given by
	\begin{equation}
		\sa=-\sc\,\bar e^0\,\bar\p_0\log\frac{\langle \lambda\,\chi_+\rangle}{\langle \lambda\,\chi_-\rangle}\,,\label{eq:twistor_connection}
	\end{equation}
where $\chi_{\pm}^{\alpha}$ are the charged Killing spinors of the SDD defined by \eqref{Kspinors}. The local nature of $\sa$ on $\PT$ is clear as it is ill-defined where $\la\lambda\,\chi_{\pm}\ra=0$, so must be understood as defined on the open patch away from these locii. It is clear that this partial connection is holomorphic, as it is given by a function on the sphere proportional to $\bar{e}^0$, so $\dbar\sa=0$ by virtue of $\bar{e}^0\wedge\bar{e}^0=0$.

Now, the defining equation \eqref{HolFDef} for the holomorphic frame associated to $E$ with this partial connection implies that a possible choice of $\sH$ is
	\be
	\sH(x,\lambda)=\left(\frac{\la \lambda\,\chi_+\ra}{\la\lambda \,\chi_-\ra}\right)^{\sc}\,.\label{eq:lambda-o_frame}
	\ee
It is easily checked that this holomorphic frame reproduces the SDD field in the gauge \eqref{eq:sdd}: using \eqref{eq:ward-gauge-field}, one obtains
	\begingroup
	\allowdisplaybreaks
	\begin{equation}
		\begin{aligned}
			-\im\lambda^\alpha A_{\alpha\dot\alpha}&=\sc\,\frac{\la \lambda\,\chi_-\ra}{\la\lambda\,\chi_+\ra}\,\lambda^\alpha\,\p_{\alpha\dot\alpha}\left(\frac{\la \lambda\,\chi_+\ra}{\la\lambda\,\chi_-\ra}\right)\\&=\sc\,\lambda^\alpha\left(\frac{\p_{\alpha\dot\alpha}\la\lambda\,\chi_+\ra}{\la \lambda\,\chi_+\ra}-\frac{\p_{\alpha\dot\alpha}\la\lambda\,\chi_-\ra}{\la\lambda\,\chi_-\ra}\right)\\&=-\im\sc\left(\lambda^\alpha\,\ta_{\alpha\dot\alpha}+\frac{\lambda^\alpha\lambda^\beta\, T^\gamma{}_{\dot\alpha}}{r^2}\left(\chi_{+\alpha}\,\chi_{-\beta}\,\chi_{+\gamma}-\chi_{-\alpha}\,\chi_{+\beta}\,\chi_{-\gamma}\right)\right)\\&=-\im\sc\,\lambda^\alpha\left(\ta_{\alpha\dot\alpha}+\frac{T_{\alpha\dot\alpha}}{r}\right)\,.
		\end{aligned}
	\end{equation}
In going from the second to the third line, we used the exterior derivatives \eqref{eq:dchi+}-\eqref{eq:dchi-}, and in the last line, we used the identity $2\chi_{+[\alpha}\chi_{-\beta]}=r\,\varepsilon_{\alpha\beta}$. Note that, as required, dependence on the open patch $\{\la\lambda\chi_{\pm}\ra\neq0\}\subset\PT$ drops out in the transform to the spacetime gauge field.  
	\endgroup

Similar computations show that the holomorphic frames associated to the SDD in Kerr-Schild gauge \eqref{eq:kerr-schild_gauge} and light-cone gauge \eqref{eq:light-cone_gauge} are given by
    \be
    \sH^\text{KS}=\left(\frac{\la\lambda\,\chi_+\ra}{\la\lambda\,\chi_-\ra}\,\frac{r}{1+\zeta\bar\zeta}\right)^{\sc}\,,\qquad\sH^{\text{LC}}=\left(\frac{\la \lambda\,\chi_+\ra\,\la\iota\,\chi_-\ra}{\la\lambda\,\chi_-\ra\,\la\iota\,\chi_+\ra}\right)^{\sc}\,,
    \ee
respectively. For instance, the light-cone gauge can be obtained by fixing the holomorphic trivialization on the twistor line $X$ as follows: we can always suppose that the holomorphic frame satisfies $\sH^\text{LC}(x,\iota)=1$, for some arbitrary $\iota_{\alpha}$. The choice of this $\iota_{\alpha}$ corresponds to a choice of complex null vector $n^{\alpha\dot\alpha}=\iota^{\alpha}\bar{\iota}^{\dot\alpha}$ on $\R^4$ which will define the light-cone gauge. One can then make the ansatz $\sH^{\text{LC}}(x,\lambda)=\exp(-g^{\text{LC}}(x,\lambda))$ for the holomorphic frame, where $g^\text{LC}$ is a function with vanishing weight in $\lambda_\alpha$ and satisfying the initial-value problem 
    \be
    \begin{cases}
    \left.\dbar\right|_Xg^\text{LC}(x,\lambda)=\left.\sa\right|_X(x,\lambda)\,,\\
    g^\text{LC}(x,\iota)=0\,.\end{cases}\label{eq:g_boundary_condition}
    \ee
The operator $\left.\dbar\right|_X=\bar{e}^0\dbar_0$ acting on $\mathcal{O}(0)$ is invertible precisely for the boundary condition \eqref{eq:g_boundary_condition}, with Green's function
    \be
    g^\text{LC}(x,\lambda)=\frac{1}{2\pi\im}\int_X\frac{\D\lambda'}{\la\lambda\,\lambda'\ra}\wedge\frac{\la\iota\,\lambda\ra}{\la\iota\,\lambda'\ra}\left.\sa\right|_X\!(x,\lambda')\,.
    \ee
This integral formula precisely reproduces the frame $\sH^\text{LC}$:
    \begingroup
    \allowdisplaybreaks
    \be
    \begin{aligned}
       \log\sH^\text{LC}(x,\lambda)&=-\frac{\sc}{2\pi\im}\int_X\frac{\D\lambda'}{\la\lambda\,\lambda'\ra}\frac{\la\iota\,\lambda\ra}{\la\iota\,\lambda'\ra}\wedge\left.\dbar\,'\right|_X\log\frac{\la\lambda'\,\chi_+\ra}{\la\lambda'\,\chi_-\ra}\\&=\frac{\sc}{2\pi\im}\int_X\D\lambda'\wedge\left.\dbar\,'\right|_X\!\left(\frac{\la\iota\,\lambda\ra}{\la\lambda\,\lambda'\ra\la\iota\,\lambda'\ra}\right)\log\frac{\la\lambda'\,\chi_+\ra}{\la\lambda'\,\chi_-\ra}\\&=\sc\int_X\D\lambda'\wedge\left(\frac{1}{\la\lambda\,\lambda'\ra}\bar\delta(\la\iota\,\lambda'\ra)+\frac{1}{\la\iota\,\lambda'\ra}\bar\delta(\la\lambda\,\lambda'\ra)\right)\la\iota\,\lambda\ra\log\frac{\la\lambda'\,\chi_+\ra}{\la\lambda'\,\chi_-\ra}\\&= \sc\log\frac{\la\lambda\,\chi_+\ra\la\iota\,\chi_-\ra}{\la\lambda\,\chi_-\ra\la\iota\,\chi_+\ra}\,.
    \end{aligned}
	\ee
    \endgroup
The second line follows from Stokes' theorem, while the third line utilizes the definition of the holomorphic delta function    
\be
\bar\delta(z)\coloneqq\frac{1}{2\pi\im}\dbar\left(\frac{1}{z}\right)=\d\bar{z}\,\delta(\mathrm{Re}\,z)\,\delta(\mathrm{Im}\,z)\,,
\ee
for any $z\in\C$.


\subsection{Quasi-momentum eigenstates}

The twistor description of the SDD allows us to encode linear, adjoint-valued, background-coupled zero-rest-mass fields in terms of twistor cohomology classes via the Penrose transform~\cite{Penrose:1969ae,Eastwood:1981jy,Ward:1990vs}. In particular, this enables us to obtain representatives of gluons written  We will focus on obtaining these background coupled fields in the gauge \eqref{eq:sdd}; the procedures for the Kerr-Schild gauge \eqref{eq:kerr-schild_gauge} and \eqref{eq:light-cone_gauge} are completely analogous. 

Recall that the two on-shell degrees of freedom of gluons in vacuum are classified by the helicity: the helicity of a gluon corresponds to whether its linearised field strength is self-dual (positive helicity) or anti-self-dual (negative helicity). On a general self-dual background, linear gluon fields still split into positive and negative helicity, but these no longer correspond to SD and ASD perturbations. Suppose we are given a linear perturbation $a_{\alpha\dot\alpha}$ to a background gauge field $A_{\alpha\dot\alpha}$, with linearised field strength $f_{ab}=f_{\alpha\beta}\epsilon_{\dot\alpha\dot\beta}+\tilde f_{\dot\alpha\dot\beta}\epsilon_{\alpha\beta}$. The linearised equations of motion for the perturbation are
	\be
	D_{\alpha\dot\alpha}f^{\alpha\beta}=0\,,\qquad D_{\alpha\dot\alpha}\tilde f^{\dot\alpha\dot\beta}+\im\,[a_{\alpha\dot\alpha},\tilde F^{\dot\alpha\dot\beta}]=0\,,
	\ee
	where $\tilde F_{\dot\alpha\dot\beta}$ is the SD field strength of $A_{\alpha\dot\alpha}$ and $D=\d-\im A$ is the gauge covariant derivative. Positive helicity gluons are then defined by $f^{\alpha\beta}=0$; that is, they are SD perturbations to the background gauge field. Conversely, it is inconsistent to set $\tilde f^{\dot\alpha\dot\beta}=0$ for a negative-helicity gluon, the obstruction being exactly the non-vanishing SD background field strength $\tilde F^{\dot\alpha\dot\beta}$. Thus, negative-helicity gluons are asymmetrically described by the background-coupled zero-rest-mass equation
	\be
D^{\alpha\dot\alpha}f_{\alpha\beta}=0\,,\label{eq:zrm_negative_gluon}
	\ee
	and will generally acquire a SD component of linearised field strength when traversing the SD background. 

The Penrose transform represents both negative and positive helicity gluons on a SD gauge field background by first cohomology classes on twistor space valued in the bundle $E\to\PT$ and twisted by $\cO(-4)$ and $\cO$, respectively~\cite{Penrose:1969ae, Eastwood:1981jy,Ward:1990vs}. In particular, we can construct twistor representatives for the particularly simple quasi-momentum eigenstate basis of gluons on the SDD background~\cite{Adamo:2023fbj}.

\medskip
	
\paragraph{Negative-helicity gluons:} The Penrose transform for background-coupled, negative helicity spin-1 fields is
	\be
	H^{0,1}_{\bar D}(\PT,\mathcal{O}(-4)\otimes\text{End}E)\cong\{a_{\alpha\dot\alpha}\text{ on }\M\,|\,D^{\alpha\dot\alpha}f_{\alpha\beta}=0\}\,,
	\ee
	where $H^{0,1}_{\bar{D}}$ denotes the Dolbeault cohomology defined with respect to $\bar D$, the (integrable) partial connection on $E\to\PT$. Given some representative $b\in H^{0,1}_{\bar D}(\PT,\mathcal{O}(-4)\otimes\text{End}E)$, the spacetime field $f_{\alpha\beta}$ can be reconstructed via the integral formula
	\be
f_{\alpha\beta}(x)=\int_X\D\lambda\wedge\lambda_\alpha\lambda_\beta\,\sH^{-1}(x,\lambda)\left.b\right|_X\sH(x,\lambda)\,,\label{eq:penrose_transform_negative_helicity}
	\ee
	where $\D\lambda\coloneqq\la\lambda\,\d\lambda\ra$ is the holomorphic $\mathcal{O}(2)$-valued $(1,0)$-form on $\P^1$ and $b|_X$ denotes the pullback of $b$ to the twistor line $X$ via the incidence relations \eqref{incidence}. Since $b$ is valued in $\mathcal{O}(-4)$ and the holomorphic frame has zero weight in $\lambda_\alpha$, the integrand is projectively well-defined and can be integrated over the line $X$. 

Let $k_{\alpha\dot\alpha}=\kappa_\alpha\tilde\kappa_{\dot\alpha}$ be a (complex) null 4-momentum, and consider representatives of the form
	\be
	b^{(e,m)}=\sT^{e}\,\int_{\C^*}\d s\,s^3\,\bar\delta^{2}(\kappa-s\,\lambda)\,\left(\frac{s^2\mu^{\dot\alpha}\,T^\alpha{}_{\dot\alpha}\,\lambda_\alpha}{\la \kappa\,o\ra^2}\right)^{m}\,\e^{\im s\,[\mu\,\tilde\kappa]}\,.\label{eq:negative_helicity_twistor_rep}
	\ee
Here the integral over the scaling parameter $s$ ensures that the representative is homogeneous of weight $-4$ on $\PT$; $2m\in\Z$ is an integer to ensure single-valuedness of the twistor
representative; and $\sT^{e}$ is a generator of the Lie algebra of the gauge group of charge $e$ with respect to the Cartan-valued SDD background: $[\sc,\sT^e]=e\sT^e$. The 2-dimensional holomorphic delta function is defined as
	\be
	\bar\delta^{2}(\kappa-s\,\lambda)\coloneqq\bigwedge_{\alpha=0,1}\dbar\left(\frac{1}{\kappa_\alpha-s\,\lambda_\alpha}\right)\,,
	\ee
where $\dbar$ is taken with respect to both $\lambda_\alpha$ and $s$. 

Since both $b^{(e,m)}$ and the partial connection $\sa$ \eqref{eq:twistor_connection} point along the $\D\hat\lambda$ direction, \eqref{eq:negative_helicity_twistor_rep} represents an element of $H^{0,1}_{\bar D}(\PT,\mathcal{O}(-4)\otimes\text{End}E)$, up to isolated singularities. Applying the integral formula \eqref{eq:penrose_transform_negative_helicity} and noticing that the incidence relations imply $\left.(\mu^{\dot\alpha}T^\alpha{}_{\dot\alpha}\lambda_\alpha)\right|_X=\langle\lambda\,\chi_+\rangle\langle\lambda\,\chi_-\rangle$, one obtains the linear field
	\be
f^{(e,m)}_{\alpha\beta}=\sT^e\,\kappa_\alpha\kappa_\beta\frac{\la\kappa\,\chi_+\ra^{m+e}\,\la\kappa\,\chi_-\ra^{m-e}}{\la\kappa\,o\ra^{2m}}\e^{\im k\cdot x}\,.\label{eq:negative_helicity_quasi_momentum_eigenstate}
	\ee
The factor $\la\kappa\,o\ra^{-2m}$ ensures that the wavefunction (on twistor space and on spacetime) has the correct scaling under little group transformations\footnote{One could worry that both the twistor representative \eqref{eq:negative_helicity_twistor_rep} and spacetime field \eqref{eq:negative_helicity_quasi_momentum_eigenstate} carry non-zero homogeneity in the constant spinor $o^{\alpha}$. This can be removed by replacing $\la\kappa\,o\ra^{-2m}$ with $\left(\frac{\la\iota\,o\ra}{\la\kappa\,\iota\ra\,\la\kappa\,o\ra}\right)^m$, which alters the expression \eqref{nhqme} below by an irrelevant factor of $z^{-m}$. We work with the given formulae to avoid the notational clutter of carrying these factors of $z$ through later calculations, but emphasize that the distinction is nothing more than overall factors.}.

It will be useful to further evaluate this expression using an affine parametrization $(z,\tilde{z})$ of the (complexified) celestial sphere of null momenta:
	\be
	\kappa^\alpha=(1,z)\,,\qquad \tilde\kappa^{\dot\alpha}=\frac{\sqrt{2}\,\omega}{1+z\tilde z}(1,\tilde z)\,,\label{eq:spinor_momenta}
	\ee
	so that \eqref{eq:negative_helicity_quasi_momentum_eigenstate} becomes
	\be\label{nhqme}
	f_{\alpha\beta}^{(e,m)}=\sT^e\,\kappa_\alpha\kappa_\beta\left(\frac{r}{1+\zeta\bar\zeta}\right)^{m}(\bar\zeta z+1)^{m+e}\,(\zeta-z)^{m-e}\,\e^{\im k\cdot x}\,.
	\ee
We recognize this parametrization as the negative helicity gluon quasi-momentum eigenstate $\phi^{(p,q)}_{\alpha\beta}$ derived in~\cite{Adamo:2023fbj} with the identifications $p\equiv m+e$ and $q\equiv m-e$. Single-valuedness of the linearised field strength requires both $p$ and $q$ to be integers, so that the charge is quantized as
	\be
	2e\in\Z\,.
	\ee
In other words, both $e$ and $m$ must be integer or half-integer valued.

In~\cite{Adamo:2023fbj}, we referred to states with $m=|e|$ as \emph{minimal} quasi-momentum eigenstates: these states are minimal in the sense that they are the states with minimal growth in $r$ at infinity which are nowhere singular on the celestial sphere. For positive and negative charge, we denote the minimal twistor representatives by
	\begingroup
	\allowdisplaybreaks
	\begin{align}
		b^{e,+}&=\sT^e\int_{\C^*}\d s\,s^3\,\bar\delta^{2}(\kappa-s\,\lambda)\left(\frac{s^2\mu^{\dot\alpha}\,T^\alpha{}_{\dot\alpha}\,\lambda_\alpha}{\la\kappa\,o\ra^2}\right)^e\,\e^{\im s\,[\mu\,\tilde\kappa]}\,,\\ b^{e,-}&=\sT^e\int_{\C^*}\d s\,s^3\,\bar\delta^{2}(\kappa-s\lambda)\left(\frac{s^2\mu^{\dot\alpha}\,T^\alpha{}_{\dot\alpha}\,\lambda_\alpha}{\la\kappa\,o\ra^2}\right)^{-e}\,\e^{\im s\,[\mu\,\tilde\kappa]}\,,
	\end{align}
	\endgroup
respectively. By contrast, ordinary momentum eigenstates correspond instead to quasi-momentum eigenstates with $m=0$; for negative charge these are singular in the direction $\la\kappa\,\chi_+\ra=0$ on the celestial sphere, while for positive charge they are singular in the $\la\kappa\,\chi_-\ra=0$ direction. Moreover, non-minimal quasi-momentum eigenstates with $m>|e|$ can be recovered from the minimal quasi-momentum eigenstates by noticing that
	\be
	\frac{(1+z\tilde z)^2}{2\omega}\,\p_{\tilde z}b^{(e,m)}=b^{(e,m+1)}\,.
	\ee
Clearly, the same relation holds for the corresponding spacetime fields.

\medskip
	
\paragraph{Positive-helicity gluons:} The Penrose transform for background-coupled, positive helicity spin-1 fields is given by
	\be
	H^{0,1}_{\bar D}(\PT,\mathcal{O}\otimes\text{End}E)\cong\{a_{\alpha\dot\alpha}\text{ on }\M\,|\,D^{\dot\alpha}_{(\alpha}a^{}_{\beta)\dot\alpha}=0\}\,,
	\ee
and in this case the linear gauge field $a_{\alpha\dot\alpha}$ itself can be constructed via an argument due to Sparling~\cite{Sparling:1990}. The twistor representative for a positive helicity quasi-momentum eigenstate is
	\be
	a^{(e,m)}=\sT^e\int_{\C^*}\frac{\d s}{s}\,\bar\delta^{2}(\kappa-s\lambda)\,\left(\frac{s^2\mu^{\dot\alpha}\,T^\alpha{}_{\dot\alpha}\,\lambda_\alpha}{\la\kappa\,o\ra^2}\right)^m\,\e^{\im s\,[\mu\,\tilde\kappa]}\,.\label{eq:positive_helicity_twistor_rep}
	\ee
	The restriction of such representative to a twistor line $X$ is $\left.\bar\p\right|_X$-exact, after a suitable dressing with the holomorphic frame:
	\be
	\begin{aligned}
\sH^{-1}a^{(e,m)}|_X\sH&=\sT^e\,\frac{\la\kappa\,\chi_+\ra^{m+e}\,\la\kappa\,\chi_-\ra^{m-e}}{\la\kappa\,o\ra^{2m}}\,\frac{\la\lambda\,\xi\ra}{\la\kappa\,\xi\ra}\,\bar\delta(\la\lambda\,\kappa\ra)\,\e^{\im k\cdot x}\\&=\left.\dbar\right|_Xj^{(e,m)}(x,\lambda)\,,
	\end{aligned}
	\ee
for
	\be
	j^{(e,m)}(x,\lambda)\coloneqq \sT^e\frac{\la\lambda\,\xi\ra}{\la\kappa\,\xi\ra\,\la\lambda\,\kappa\ra}\frac{\la\kappa\,\chi_+\ra^{m+e}\,\la\kappa\,\chi_-\ra^{m-e}}{\la\kappa\,o\ra^{2m}}\,\e^{\im k\cdot x}\,.
	\ee
Here, $\xi_{\alpha}$ is an arbitrary reference spinor whose choice reflects the residual linearised gauge freedom in the corresponding spacetime potential $a_{\alpha\dot\alpha}^{(e,m)}$. This linear gauge field is recovered by noticing that $\lambda^\alpha D_{\alpha\dot\alpha}j^{(e,m)}$ is a holomorphic function of $\lambda_\alpha$ of weight +1, so Liouville's theorem gives
	\be
	\lambda^\alpha\p_{\alpha\dot\alpha}j^{(e,m)}=-\im\lambda^\alpha \,a^{(e,m)}_{\alpha\dot\alpha}\,,\qquad a^{(e,m)}_{\alpha\dot\alpha}=\sT^e\,\frac{\xi_\alpha \,\tilde K_{\dot\alpha}^{(e,m)}(x)}{\la\kappa\,\xi\ra}\,\frac{\la\kappa\,\chi_+\ra^{m+e}\,\la\kappa\,\chi_-\ra^{m-e}}{\la\kappa\,o\ra^{2m}}\,\e^{\im k\cdot x}\,.
	\ee
In this expression,
	\be
	\tilde K_{\dot\alpha}^{(e,m)}(x):=\tilde\kappa_{\dot\alpha}+\im(m+e)\,\frac{\chi_+^\gamma\, T_{\gamma\dot\alpha}}{r\,\la\kappa\,\chi_+\ra}-\im(m-e)\,\frac{\chi_-^\gamma\, T_{\gamma\dot\alpha}}{r\,\la\kappa\,\chi_-\ra}\,,
	\ee
is the \emph{background-dressed} dotted momentum spinor of the gluon perturbation.

Minimal quasi-momentum eigenstates are defined in exactly the same way as for the negative helicity case, namely $m=|e|$. These correspond to the twistor representatives
	\begingroup
	\allowdisplaybreaks
	\begin{align}
		a^{e,+}&=\sT^e\int_{\C^*}\frac{\d s}{s}\,\bar\delta^{2}(\kappa-s\lambda)\left(\frac{s^2\mu^{\dot\alpha}\,T^\alpha{}_{\dot\alpha}\,\lambda_\alpha}{\la\kappa\,o\ra^2}\right)^e\,\e^{\im s\,[\mu\,\tilde\kappa]}\,,\\ a^{e,-}&=\sT^e\int_{\C^*}\frac{\d s}{s}\,\bar\delta^{2}(\kappa-s\lambda)\left(\frac{s^2\mu^{\dot\alpha}\,T^\alpha{}_{\dot\alpha}\,\lambda_\alpha}{\la\kappa\,o\ra^2}\right)^{-e}\,\e^{\im s\,[\mu\,\tilde\kappa]}\,,
	\end{align}
	\endgroup
	for positive and negative charge, respectively. Non-minimal quasi-momentum eigenstates can be again constructed by recursively applying the relation
	\be
	\frac{(1+z\tilde z)^2}{2\omega}\,\p_{\tilde z}a^{(e,m)}=a^{(e,m+1)}\,,
	\ee
which also holds for the wavefunctions on spacetime.


\section{MHV scattering around the self-dual dyon}\label{sec:mhv}

Armed with the twistor description of the SDD and background-coupled quasi-momentum eigenstate gluon wavefunctions, we now turn to the computation of scattering amplitudes. In this section, we first review the spacetime generating functional for the maximal helicity violating (MHV) gluon amplitude on the SDD background. We then proceed to lift this generating functional to twistor space, where the perturbative expansion around the SDD background is particularly straightforward and leads to an \emph{all-multiplicity} formula for the MHV amplitude. This is a generalization of the Parke-Taylor formula~\cite{Parke:1986gb}, to which it reduces when each external gluon has vanishing quantum numbers $m$ and $e$. In general, our formula takes the form of an integral over a single spatial point, and we discuss special cases where this integral can be evaluated explicitly, including at low multiplicity and whenever there is a single negatively-charged external gluon.


\subsection{The MHV generating functional}

In the perturbiner approach to scattering amplitudes \cite{Arefeva:1974jv,Abbott:1983zw,Jevicki:1987ax,Selivanov:1999as}, tree-level amplitudes are computed as multi-linear pieces of the classical action, evaluated on recursively constructed solutions of the non-linear equations of motion with appropriate boundary conditions. To compute gluon tree amplitudes in flat space, we can use the Chalmers-Siegel formulation~\cite{Chalmers:1996rq} of Yang-Mills theory as a perturbative expansion around the self-dual sector. This is given by the first-order action
	\be\label{Ch-Siact}
	S[A,B]=\int_\M\d^4x\,\tr\left( B_{\alpha\beta}\,F^{\alpha\beta}\right)-\frac{\rg^2}{2}\int_\M\d^4x\,\tr\left( B_{\alpha\beta}\,B^{\alpha\beta}\right)\,,
	\ee
where $F^{\alpha\beta}$ is the anti-self-dual (ASD) field strength of the Yang-Mills gauge field $A$, $B_{\alpha\beta}$ is an adjoint-valued ASD 2-form , $\rg$ is the gauge coupling and the traces are over the Lie algebra of the gauge group. The equations of motion arising from this action are
	\be
	D^{\alpha\dot\alpha}B_{\alpha\beta}=0\,,\qquad F_{\alpha\beta}=\rg^2\,B_{\alpha\beta}\,,
	\ee
so that integrating out the 2-form $B$ one obtains an action that differs from the standard Yang-Mills action only by a topological term. Thus, the Chalmers-Siegel and Yang-Mills actions are perturbatively equivalent (with asymptotically flat boundary conditions), and \eqref{Ch-Siact} will compute the correct tree-level gluon scattering amplitudes. 
	
Suppose now that we are given a self-dual background $A_{\alpha\dot\alpha}$ and a collection of $n$ gluons propagating on this background. Let the $r^\text{th}$ and $s^{\text{th}}$ gluons have negative helicity, whilst the other $n-2$ gluons have positive helicity and are described by the linearized gauge fields $\{a^{(i)}_{\alpha\dot\alpha}\}_{1\leq i\leq n}^{i\neq r,s}$. Since positive helicity gluons represent self-dual perturbations to the background, they can be viewed as a defining a coherent state and, together with the background $A_{\alpha\dot\alpha}$ itself, they define a \emph{new} self-dual field $\mathcal{A}_{\alpha\dot\alpha}$, which in general depends non-linearly on each of the $a^{(i)}_{\alpha\dot\alpha}$. The two negative helicity gluons are described by linearized ASD field strengths $f^{(r)}_{\alpha\beta}$ and $f^{(s)}_{\alpha\beta}$ in this background. 

In the perturbiner framework, the generating functional for the gluon MHV amplitude is precisely the 2-point function of the negative-helicity gluons in the deformed SD background $\mathcal{A}$~\cite{Mason:2009afn,Adamo:2020yzi}
	\be
	\mathcal{G}(r,s)=\frac{1}{2\,\rg^2}\int_\M\d^4x\,\tr\left( f^{(r)}_{\alpha\beta}\,f^{(s)\,\alpha\beta}\right)\,.\label{eq:mhv_generating_functional}
	\ee
This means that the $n$-point MHV amplitude on the SD background $A$ is precisely given by the term in \eqref{eq:mhv_generating_functional} which is linear in each of the external positive helicity gluons $\{ a^{(i)}_{\alpha\dot\alpha}\}_{1\leq i\leq n}^{i\neq r,s}$.

\medskip
	
\paragraph{Uplift to twistor space:} The computation of such a linear piece of the action using spacetime methods is rather difficult, mainly due to the complication of expanding the generating functional \eqref{eq:mhv_generating_functional}. However, this computation becomes straightforward when translated to twistor space. Suppose that $A_{\alpha\dot\alpha}$ is described on twistor space via the Ward correspondence in terms of the partial connection $\bar D=\dbar+\sa$ on the holomorphic vector bundle $E\to\PT$. By the Penrose tranform, the negative helicity gluons are described by twistor representatives $b^{(r)},b^{(s)}\in H^{0,1}_{\bar D}(\PT,\mathcal{O}(-4)\otimes\text{End}E)$, and the $i^{\text{th}}$ positive helicity gluon is described by a twistor representative $a^{(i)}\in H^{0,1}_{\bar D}(\PT,\mathcal{O}\otimes\text{End}E)$. Without loss of generality, we can assume\footnote{This assumption corresponds to writing the representatives of the cohomology class in a `radiative gauge', defined in terms of the asymptotic characteristic data on spacetime~\cite{Penrose:1980yx,Mason:1986,Adamo:2021dfg}.} that the $(0,1)$-form components of all of these gluon representatives are proportional to  $\D\hat\lambda$. 

A deformed partial connection on $E$ can now be defined by $\tilde{D}:=\bar{D}+a$, where
	\be
	a:=\sum_{\substack{i=1\\i\neq r,s}}^n \rg\,\veps_i\,a^{(i)}\,,
	\ee
for $\veps_i$ formal parameters. Under our assumptions, $\tilde{D}^2=0$ so the deformed partial connection also defines a holomorphic structure on $E$. Thus, $b^{(r)}$, $b^{(s)}$ are also representatives of classes in $H^{0,1}_{\tilde{D}}(\PT,\mathcal{O}(-4)\otimes\text{End}E)$. Let $\tilde\sH$ and $\sH$ denote the holomorphic frames associated with the partial connections $\tilde{D}=\bar{D}+a$ and $\bar{D}$, respectively.

The uplift to twistor space of the generating functional \eqref{eq:mhv_generating_functional} is now readily obtained via the Penrose transform \eqref{eq:penrose_transform_negative_helicity} for negative-helicity fields:
	\be
	\mathcal{G}(r,s)=\frac{1}{2\,\rg^2}\int\d^4x\,\D\lambda_1\,\D\lambda_2\,\la\lambda_1\,\lambda_2\ra^2\,\tr\left(\tilde\sH_1^{-1}\,b^{(r)}_1\,\tilde\sH_1\,\tilde\sH_2^{-1}\,b^{(s)}_2\,\tilde\sH_2\right)\,.
	\ee
The integrals here are over two copies of the twistor line $X$, with homogeneous coordinates $\lambda_{1\,\alpha}$ and $\lambda_{2\,\alpha}$ respectively, and there is an additional integration over the moduli of such lines, namely an integral over spacetime. We use the notation $\tilde\sH_1$ and $b_1^{(r)}$ for $\tilde\sH(x,\lambda_1)$ and $\left.b^{(r)}\right|_X(x,\lambda_1)$, and similarly for $\tilde{\sH}_2$ and $b_2^{(s)}$. The generating functional can be rewritten as
	\be
	\mathcal{G}(r,s)=\frac{2\pi^2}{\rg^2}\int\d^4x\,\D\lambda_1\,\D\lambda_2\,\la\lambda_1\,\lambda_2\ra^4\,\tr\left(b^{(r)}_1\,\tilde\sK_X(\lambda_1,\lambda_2)\,b^{(s)}_2\,\tilde\sK_X(\lambda_2,\lambda_1)\right)\,,\label{eq:twistor_mhv_generating_functional}
	\ee
	where we introduced the function
	\be
\tilde\sK_X(\lambda,\lambda')\coloneqq\frac{\tilde\sH(x,\lambda)\,\tilde\sH^{-1}(x,\lambda')}{2\pi\im\,\la\lambda\,\lambda'\ra}\,.
	\ee
Given the defining equation for the frame $\tilde\sH$ (i.e., $\tilde{D}|_{X}\tilde{\sH}=0$), it is clear that $\tilde \sK_X$ is the Green's function for the operator $\tilde{D}|_X$ acting on sections of $\mathcal{O}(-1)\to X$. By the same argument, the Green's function for the undeformed operator $\left.\bar\D\right|_X=\left.\dbar\right|_X+\left.\sa\right|_X$ is
\be	\sK_X(\lambda,\lambda'):=\frac{\sH(x,\lambda)\,\sH^{-1}(x,\lambda')}{2\pi\im\,\la\lambda\,\lambda'\ra}\,.
	\ee
Crucially, this representation for the MHV generating functional can now be straightforwardly expanded in $a$. 


\subsection{The MHV amplitude}

The MHV amplitude can now be extracted from \eqref{eq:twistor_mhv_generating_functional} by obtaining a relation between the Green's functions $\tilde\sK_X$ and $\sK_X$. The defining equation for $\tilde\sK_X$ is
	\be
\left.\bar{D}\right|_X\tilde\sK_X(\lambda,\lambda')+\left.a\right|_X\tilde\sK_X(\lambda,\lambda')=\bar\delta(\la\lambda\,\lambda'\ra)\,,
	\ee
which can be integrated to
	\be
\tilde\sK_X(\lambda,\lambda')=\sK_X(\lambda,\lambda')-\int_X\D\lambda''\,\sK_X(\lambda,\lambda'')\left.a\right|_X(\lambda'')\,\tilde\sK(\lambda'',\lambda')\,,
	\ee
in terms of the Green's function $\sK_X$ for $\bar{D}|_{X}$. This integral formula can be iterated to obtain the Born series
\be\label{Bornser}	\tilde\sK_X(\lambda,\lambda')=\sum_{n=0}^\infty\left(\frac{-1}{2\pi\im}\right)^n\int_{X^n}\D\lambda_1\cdots\D\lambda_n\frac{\sH(x,\lambda)\sH_1^{-1}a_1\sH_1\cdots\sH_n^{-1}a_n\sH_n\sH^{-1}(x,\lambda')}{\la\lambda\,\lambda_1\ra\,\la\lambda_1\,\lambda_2\ra\cdots\la\lambda_{n-1}\,\lambda_n\ra\,\la\lambda_n\,\lambda'\ra}\,.
	\ee
Here we denote $\sH_i=\sH(x,\lambda_i)$ and $a_i=\left.a\right|_X(x,\lambda_i)$. Moreover, the $n=0$ term in this series is understood to be $\sK_X(\lambda,\lambda')$. 

The MHV amplitude is now given as the coefficient of $\veps_1\cdots\veps_n$ in the generating functional, expanded using \eqref{Bornser}:
	\begin{multline}
		\mathcal{A}^{\mathrm{MHV}}_n=\rg^{n-2}\int_{\M\times X^n}\d^{4}x\,\D\lambda_1\cdots\D\lambda_n\,\la\lambda_r\,\lambda_s\ra^4 \\
  \times\frac{\tr\left(\sH_1^{-1}a^{(1)}_1\sH_1\cdots\sH_r^{-1}b^{(r)}_r\sH_r\cdots\sH^{-1}_sb^{(s)}_s\sH_s\cdots\sH_n^{-1}a^{(n)}_n\sH_n\right)}{\la\lambda_1\,\lambda_2\ra\,\la\lambda_2\,\lambda_3\ra\cdots\la\lambda_{n-1}\,\lambda_n\ra\,\la\lambda_n\,\lambda_1\ra}\,,
	\end{multline}
where we have written only one of the colour-orderings arising from the perturbative expansion, namely the $(123\cdots n-1\,n)$ ordering. The full amplitude is given by a sum over all non-cyclic permutations of this expression, but knowledge of a single colour ordering suffices.

We now evaluate this formula on the SDD background -- that is, for $\sa$ given by \eqref{eq:twistor_connection} -- with each external gluon represented by a quasi-momentum eigenstate. Denote the momentum and quantum numbers of the $i^\text{th}$ gluon as $\{k_i^{\alpha\dot\alpha}=\kappa_i^\alpha\tilde\kappa_i^{\dot\alpha},e_i,m_i\}$, recalling that $e_i$ is the charge of the gluon with respect to the SDD background while $m_i$ controls the radial growth of the quasi-momentum eigenstate wavefunction. In this case, the $\P^1$ integrals localize against the holomorphic delta functions in either \eqref{eq:negative_helicity_quasi_momentum_eigenstate} or \eqref{eq:positive_helicity_twistor_rep}, so the colour-ordered, tree-level, $n$-point MHV amplitude reads
	\be
	\boxed{\hspace{1em}\begin{aligned}\mathcal{A}^{\mathrm{MHV}}_n(\{\kappa_i^\alpha,\tilde\kappa_i^{\dot\alpha},&e_i,m_i\})=2\pi\, \rg^{n-2}\,\delta(\omega)\,\delta(e)\,\frac{\la r\,s\ra^4}{\la 1\,2\ra\cdots\la n\,1\ra}\times\\&\times\int\d^3\vec x\,\e^{\im \vec k\cdot\vec x}\,\prod_{a=1}^n\left(\frac{r}{1+\zeta\bar\zeta}\right)^{m_a}(\bar\zeta z_a+1)^{e_a+m_a}\,(\zeta-z_a)^{e_a-m_a}\,,\end{aligned}\hspace{1em}}\label{eq:mhv_amplitude}
	\ee
where the total energy, momentum and charge are defined as
	\be
	\omega\coloneqq\sum_{i=1}^n\omega_i\,,\qquad \vec k\coloneqq\sum_{i=1}^n\vec k_i\,,\qquad e\coloneqq\sum_{i=1}^ne_i\,,
	\ee
respectively.    

This formula is our main result. The $\delta$ function in the energy arises from the time integral and from the fact that quasi-momentum eigenstates depend on the time coordinate purely through the factor $\e^{-\im\omega_ix^0}$, as a consequence of the time-translation symmetry of the underlying background. The $\delta$ function in the charge arises from the trace over the gauge algebra indices, and ensures that the amplitude is invariant under gauge transformations of the SDD background. The presence of the background leads to the integral over the single spatial slice appearing in the second line, which replaces the three spatial momentum conserving delta functions familiar from scattering in trivial backgrounds. It is remarkable that the formula features only a single spatial integral regardless of the number of external positive helicity gluons.


\subsection{Explicit examples}
	
While the formula \eqref{eq:mhv_amplitude} is already remarkably simple, for special configurations of the quantum numbers of the external gluons, it is possible to further simplify the formula by carrying out the spatial integrals explicitly. For clarity, some technical details of the computation are collected in Appendix~\ref{app:integration}. We will now restrict our attention to the scattering of \emph{minimal} quasi-momentum eigenstates, recalling that the amplitude for more general states can be obtained by acting with simple differential operators in the external momenta.

Let $A_n$ be the $n$-point MHV amplitude for minimal quasi-momentum eigenstates, define $\mathfrak n\coloneqq\{1,\ldots, n\}$ and introduce the subsets $\mathfrak n_\pm\subseteq\mathfrak n$ corresponding to positive- and negative-charge gluons. The amplitude can be compactly written as
	\begin{multline}
		A_n=2\pi\, \rg^{n-2}\,\delta(\omega)\,\delta(e)\,\frac{\la r\,s\ra^4}{\la 1\,2\ra\cdots\la n\,1\ra}\\\times\int\d^3\vec x\,\e^{\im \vec k\cdot\vec x}\prod_{a\,\in\,\mathfrak n_+}\left(\frac{r}{1+\zeta\bar\zeta}\right)^{e_a}(\bar\zeta z_a+1)^{2e_a}\prod_{b\,\in\,\mathfrak n_-}\left(\frac{r}{1+\zeta\bar\zeta}\right)^{-e_b}(\zeta-z_b)^{-2e_b}\,.\label{eq:minimal_amplitude}
	\end{multline}
First, let us consider the 2-point amplitude, with no positive helicity extenal gluons. Without loss of generality, suppose gluon 2 has positive charge $e_2>0$, so that on the support of charge conservation
	\be
A_2=-2\pi\,\delta(\omega)\,\delta(e)\,\la1\,2\ra^2\int\d^3\vec x\,\e^{\im \vec k\cdot\vec x}\left[\frac{r(\zeta-z_1)(\bar\zeta z_2+1)}{1+\zeta\bar\zeta}\right]^{2e_2}\,.\label{eq:integrated_2point}
	\ee
Up to some irrelevant overall numerical factors, this is precisely the 2-point amplitude that we presented in \cite{Adamo:2023fbj}. The integral can be done with the help of a master Gaussian integral and the integrated form of the amplitude reads
	\be
	A_2=-16\pi^2\, e_2\,(2e_2)!\,\la1\,2\ra^2\,\frac{z_{12}\,(-\im\omega_1\,z_{12})^{2e_2-1}}{|\vec k_1+\vec k_2|^{2+4e_2}}\,,
	\ee
for $z_{12}\coloneqq z_1-z_2$.
	
it is also possible to obtain the integrated MHV amplitude explicitly at 3-points. There are two relevant cases, depending on whether there are two gluons with positive or negative charge, respectively. Let us suppose that the first gluon always has charge with opposite sign to the charge of the second and third gluon, so that the minimal MHV amplitude is
	\begin{multline}
		A_3=2\pi\, \rg\,\delta(\omega)\,\delta(e)\,\frac{\la r\,s\ra^4}{\la 1\,2\ra\,\la2\,3\ra\,\la3\,1\ra}\\\times\int\d^3\vec x\,\e^{\im \vec k\cdot\vec x}\left[\frac{r(\zeta-z_1)(\bar\zeta z_2+1)}{1+\zeta\bar\zeta}\right]^{2e_2}\left[\frac{r(\zeta-z_1)(\bar \zeta z_3+1)}{1+\zeta\bar\zeta}\right]^{2e_3}\,,
	\end{multline}
	or
	\begin{multline}
		A_3=2\pi\, \rg\,\delta(\omega)\,\delta(e)\,\frac{\la r\,s\ra^4}{\la 1\,2\ra\la2\,3\ra\la3\,1\ra}\\\times\int\d^3\vec x\,\e^{\im \vec k\cdot\vec x}\left[\frac{r(\zeta-z_2)(\bar\zeta z_1+1)}{1+\zeta\bar\zeta}\right]^{-2e_2}\left[\frac{r(\zeta-z_3)(\bar \zeta z_1+1)}{1+\zeta\bar\zeta}\right]^{-2e_3}\,,
	\end{multline}
	respectively. In both cases, the integrated amplitude reads
	\begin{multline}
		A_3=8\pi^2\im\, \rg\,\delta(\omega)\,\delta(e)\,\frac{\la r\,s\ra^4}{\la 1\,2\ra\,\la2\,3\ra\,\la3\,1\ra}\,(-1)^{2e_1}\,(2|e_1|)!\\\times\frac{(\im\alpha_{21;3})^{2|e_2|}\,(\im \alpha_{31;3})^{2|e_3|}}{|\vec k_1+\vec k_2+\vec k_3|^{2+4|e_1|}}\left(\frac{2|e_2|\,z_{21}}{ \alpha_{21;3}}+\frac{2|e_3|\,z_{31}}{ \alpha_{31;3}}\right)\,,\label{eq:integrated_3point}
	\end{multline}
	where
	\be
	\begin{aligned}\alpha_{21;3}&=\omega_{12}z_{12}+\la 1|T|3]\la3\,2\ra+\la 2|T|3]\la3\,1\ra\,,\\\alpha_{31;3}&=\omega_{13}z_{13}+\la 1|T|2]\la2\,3\ra+\la 3|T|2]\la2\,1\ra\,,
    \end{aligned}
	\ee
for $\omega_{ij}:=\omega_i-\omega_j$.	
	
More generally, it is possible to obtain a similar result whenever there is just a single negative-charge gluon or a single positive-charge gluon. If the first gluon has charge with opposite sign to the charge of the other $n-1$ gluons, the minimal MHV amplitude is
	\begin{multline}
		A_n=8\pi^2\,\im\, \rg^{n-2}\,\delta(\omega)\,\delta(e)\,\frac{\la r\,s\ra^4}{\la 1\,2\ra\cdots\la n\,1\ra}(-1)^{2e_1}\,(2|e_1|)!\\\times\frac{(\im\alpha_{21;n})^{2|e_2|}\cdots(\im\alpha_{n1;n})^{2|e_n|}}{|\vec k_1+\ldots+\vec k_n|^{2+4|e_1|}}\,\sum_{a=2}^n\frac{2|e_a|\,z_{a1}}{\alpha_{a1;n}}\,,\label{eq:integrated_npoint}
	\end{multline}
	where we define the kinematic quantity
	\be
	\alpha_{ij;n}:=\sum_{l=1}^n\la i|T|l]\,\la l\,j\ra+(i\leftrightarrow j)\,.
	\ee
Presumably, the integration can be carried out for generic charge configurations, but the final expression for the amplitude is already fairly involved in this`single negative charge' case. We leave the explicit evaluation of the integral in the general case to future work.

	
\section{Holomorphic collinear limits on the self-dual dyon}\label{sec:ope}

Armed with the explicit formula \eqref{eq:mhv_amplitude}, we can study infrared (IR) features of gluon scattering on the SDD. In particular, the MHV sector encodes universal information about holomorphic collinear splitting functions, which are equivalent to holomorphic OPE coefficients in the context of celestial holography~\cite{Fan:2019emx,Pate:2019lpp,Banerjee:2020kaa,Fotopoulos:2020bqj,Banerjee:2020zlg,Banerjee:2020vnt,Himwich:2021dau} and provide important bottom-up data for potential holographic descriptions of scattering. Furthermore, holomorphic chiral symmetry algebras associated with the self-dual sector are encoded through these holomorphic collinear limits~\cite{Guevara:2021abz,Strominger:2021lvk,Jiang:2021ovh,Adamo:2021lrv,Costello:2022wso,Bu:2022dis}. In general, it is known that these celestial OPE coefficients and the associated chiral algebras are deformed in the presence of self-dual background fields or non-commutativity~\cite{Bu:2022iak,Monteiro:2022lwm,Monteiro:2022xwq,Costello:2022jpg,Costello:2023hmi,Garner:2023izn,Bittleston:2023bzp}, although they are preserved in any self-dual radiative background~\cite{Adamo:2023zeh}. We find that the SDD background preserves the holomorphic celestial OPE in a trivial background for all helicity configurations, although this seems to be due primarily to the fact that the background is Cartan-valued. 
	
To calculate the holomorphic collinear limits, it is useful to express the holomorphic frames in terms of spinor variables and to not integrate over the time direction, so that the appropriate presentation of the MHV amplitude is
	\begin{multline}
		\mathcal{A}_n^\text{MHV}=\rg^{n-2}\,\frac{\la r\,s\ra^4}{\la1\,2\ra\cdots\la n\,1\ra}\,\tr(\sT^{e_1}\ldots \sT^{e_n})\\\times\int\d^4x\,\prod_{a=1}^n\e^{\im k_a\cdot x}\,\frac{\la a\,\chi_+\ra^{m_a+e_a}\,\la a\,\chi_-\ra^{m_a-e_a}}{\la a\,o\ra^{2m_a}}+\text{perms.}\,,
	\end{multline}
where the sum is over all non-cyclic permutations of the external gluons. In particular, this is the full, non-colour-ordered amplitude. Now, consider the case where the $i^\text{th}$ and $j^{\text{th}}$ gluons become holomorphically collinear; this limit is implemented by writing
	\be
	k_i+k_j=p+\eps^2\, q\,,
	\ee
	where $p^{\alpha\dot\alpha}=\kappa_p^\alpha\tilde\kappa_p^{\dot\alpha}$ is the on-shell collinear momentum, $q^{\alpha\dot\alpha}=\xi^\alpha\,\tilde\xi^{\dot\alpha}$ is a reference null vector, and the holomorphic limit corresponds $\la i\,j\ra\sim\eps\to0$ whilst $[i\,j]$ is held fixed. This means that we take
	\be
	\kappa_i^\alpha=\frac{\la\xi\,i\ra}{\la\xi\,p\ra}\,\kappa_p^\alpha+O(\eps)\,,\qquad\kappa_j^\alpha=\frac{\la\xi\,j\ra}{\la\xi\,p\ra}\,\kappa_p^\alpha+O(\eps)\,,
	\ee
and the holomorphic splitting functions in the SDD background are extracted as the coefficients of the $\la i\,j\ra^{-1}$ singularity in the MHV amplitude. The calculation entails considering two cases: where the collinear gluons have the same or opposite helicities. We now consider each case in turn.

\medskip
	
\paragraph{Same helicity:} Let us start with two positive helicity collinear gluons. The only colour orderings that can give a singular contribution in the holomorphic collinear limit are those where the $i^\text{th}$ and $j^\text{th}$ gluons are adjacent, so that
	\begin{multline}
		\mathcal{A}_n^\text{MHV}=\rg^{n-2}\left[\frac{\la r\,s\ra^4}{\la 1\,2\ra\cdots\la i-1\,i\ra\la i\,j\ra\la j\,j+1\ra\cdots \la n\,1\ra}\tr (\sT^{e_1}\cdots \sT^{e_i}\sT^{e_j}\cdots\sT^{e_n})+\right.\\\left.+\frac{\la r\,s\ra^4}{\la 1\,2\ra\cdots\la i-1\,j\ra\la j\,i\ra\la i\,j+1\ra\cdots \la n\,1\ra}\tr (\sT^{e_1}\cdots \sT^{e_i}\sT^{e_j}\cdots\sT^{e_n})\right]\\\times\int\d^{4}x\,\prod_{a=1}^n\e^{\im k_a\cdot x}\,\frac{\la a\,\chi_+\ra^{m_a+e_a}\,\la a\,\chi_-\ra^{m_a-e_a}}{\la a\,o\ra^{2m_a}}+O(\eps)\,.
	\end{multline}
	The leading order singularity is then simply
	\begin{multline}
		\mathcal{A}_n^\text{MHV}=\frac{\rg\,\la\xi \,p\ra^2}{\la i\,j\ra\,\la \xi\,i\ra\,\la\xi\,j\ra}\,\frac{\rg^{n-3}\,\la r\,s\ra^4}{\la 1\,2\ra\cdots\la i-1\,p\ra\la p\,j+1\ra\cdots\la n\,1\ra}\,\tr(\sT^{e_1}\cdots [\sT^{e_i},\sT^{e_j}]\cdots \sT^{e_n})\\\times\int\d^4x\,\e^{\im p\cdot x}\frac{\la p\,\chi_+\ra^{m_p+e_p}\,\la p\,\chi_-\ra^{m_p-e_p}}{\la p\,o\ra^{2m_p}}\prod_{\substack{a=1\\a\neq i,j}}^n\e^{\im k_a\cdot x}\frac{\la a\,\chi_+\ra^{m_a+e_a}\,\la a\,\chi_-\ra^{m_a-e_a}}{\la a\,o\ra^{2m_a}}\,,
	\end{multline}
	where $e_p=e_i+e_j$ and $m_p=m_i+m_j$. Note that $[\sT^{e_i},\sT^{e_j}]$ has precisely charge $e_p$, as implied by the Jacobi identity. The holomorphic collinear limit is thus
	\begin{multline}
		\mathcal{A}_n^{\text{MHV}}(1^+,\ldots,i^+,j^+,\ldots,n^+)\to\frac{1}{\la i\,j\ra}\,\text{Split}(i^{+,e_i,m_i},j^{+,e_i,m_i}\to p^{+,e_p,m_p})\times\\\times\mathcal{A}_{n-1}^\text{MHV}(1^+,\ldots,p^+,\ldots,n^+)\,,
	\end{multline}
	with splitting function
	\be
	\text{Split}(i^{+,e_i,m_i},j^{+,e_i,m_i}\to p^{+,e_p,m_p})=\frac{\rg\,\la\xi\,p\ra^2}{\la\xi\,i\ra\,\la\xi\,j\ra}\,\delta_{m_p,m_i+m_j}\,\delta_{e_p,e_i+e_j}\,.
	\ee
	This is the holomorphic splitting function that arises for the same-helicity collinear limit in a trivial background~\cite{Altarelli:1977zs,Mangano:1990by,Birthwright:2005ak}, so we see that the SDD leaves the splitting function invariant. After a Mellin transform, this implies that the associated celestial OPE coefficient and chiral algebra are similarly un-altered.
 
\medskip
	
\paragraph{Opposite helicity:} In the case where the two collinear gluons have opposite helicity, the computation proceeds along the same lines as before. Suppose we take the collinear limit between the $r^\text{th}$ and $i^\text{th}$ gluon, so that the leading singularity in the $n$-point amplitude is
	\begin{multline}
		\mathcal{A}_n^\text{MHV}=\frac{\rg\,\la\xi\,r\ra^3}{\la r\,i\ra\,\la\xi\,i\ra\,\la\xi\,p\ra^2}\,\frac{\rg^{n-3}\,\la p\,s\ra^4}{\la 1\,2\ra\cdots\la r-1\,p\ra\,\la p\,i+1\ra\ldots\la n\,1\ra}\tr(\sT^{e_1}\cdots[\sT^{e_r},\sT^{e_i}]\cdots\sT^{e_n})\\\times\int\d^4x\,\e^{\im p\cdot x}\frac{\la p\,\chi_+\ra^{m_p+e_p}\la p\,\chi_-\ra^{m_p-e_p}}{\la p\,o\ra^{2m_P}}\prod_{\substack{a=1\\a\neq i,r}}^n\e^{\im k_a\cdot x}\frac{\la a\,\chi_+\ra^{m_a+e_a}\la a\,\chi_-\ra^{m_a-e_a}}{\la a\,o\ra^{2m_a}}\,,
	\end{multline}
	so that the splitting function is now
	\be
	\text{Split}(i^{+,e_i,m_i},r^{-,e_r,m_r}\to p^{-,e_p,m_p})=\frac{\rg\,\la\xi\,r\ra^3}{\la\xi\,i\ra\,\la\xi\,p\ra^2}\,\delta_{m_p,m_r+m_i}\,\delta_{e_p,e_r+e_i}\,.
	\ee
	Again, the splitting function is un-deformed by the presence of the SDD background~\cite{Altarelli:1977zs,Mangano:1990by,Birthwright:2005ak}.
	
		
	\section{Conclusion}\label{sec:concl}
 
In this paper, we used twistor theory to compute multi-gluon MHV amplitudes in the background of a self-dual dyon gauge field. This relied on a twistor description of the SDD itself as well as the gluon quasi-momentum eigenstates derived using charged Killing spinors in~\cite{Adamo:2023fbj}. In general, our formula for the MHV amplitude \eqref{eq:mhv_amplitude} is left as an integral over a single spatial slice, which replaces the usual conservation of spatial 3-momentum, reflecting the fact that the background breaks translation invariance. We showed that this integral can be performed explicitly in low-multiplicity cases (such as two- and three-points), as well as at arbitrary multiplicity when there is a single negatively (or a single positively) charged gluon. Our formula enabled us to determine the holomorphic collinear splitting functions for gluons in the SDD background, which were shown to be equivalent to those of a trivial background.

There are a number of avenues for further exploration. While we studied the leading holomorphic collinear limits of the MHV amplitude, it should also be able to obtain an \emph{all-order} holomorphic collinear expansion of our amplitude using methods similar to flat space~\cite{Adamo:2022wjo,Ren:2023trv}. While the leading splitting functions were un-deformed, one expects that such an all-orders collinear expansion would be sensistive to the SDD background through the sub-leading terms, as shown for MHV scattering on SD radiative backgrounds~\cite{Adamo:2023zeh}. 

The MHV amplitude is, of course, just one helicity sector of the full tree-level gluon S-matrix on the SDD background. Taking inspiration from similar formulae for all gluon amplitudes and form factors in trivial~\cite{Witten:2003nn,Roiban:2004yf,Koster:2016loo,He:2016jdg,Koster:2016fna} and self-dual radiative backgrounds~\cite{Adamo:2020syc,Bogna:2023bbd}, we can make a natural conjecture for the $n$-point, N$^k$MHV amplitude on the SDD background in terms of an integral over the moduli space of degree $k+1$ holomorphic maps from the Riemann sphere to twistor space. In particular, let $\sigma^{\mathbf{a}}=(\sigma^{\mathbf{0}},\sigma^{\mathbf{1}})$ be homogeneous holomorphic coordinates on $\P^1$; a degree $k+1$ holomorphic map $Z:\P^1\to\PT$ can be parametrized by
\be\label{hmap}
Z^{A}(\sigma)=U^{A}_{\mathbf{a}_1\cdots\mathbf{a}_{k+1}}\,\sigma^{\mathbf{a}_1}\cdots\sigma^{\mathbf{a}_{k+1}}\,,
\ee
with an overall GL$(2,\C)$ redundancy composed of projective rescalings and M\"obius transformations on the sphere. Let $\tilde{\mathbf{g}}\subset\{1,\ldots,n\}$ be the set of $k+2$ negative helicity gluons in the N$^k$MHV scattering process, with $\mathbf{g}$ the complementary set of positive helicity gluons.

We can then make the following conjecture for the colour-ordered N$^k$MHV partial amplitude for gluon scattering in the SDD background:
\be\label{NkMHV}
\cA_{n,k}=\int\frac{\d^{4(k+2)}U}{\mathrm{vol}\,\GL(2,\C)}\prod_{\substack{i<j \\ i,j\in\tilde{\mathbf{g}}}}(i\,j)^4\,\prod_{i=1}^{n}\frac{\D\sigma_i}{(i\,i+1)}\,\prod_{j\in\mathbf{g}}a_j(Z(\sigma_j))\,\e^{e_j\,g(\sigma_j)}\prod_{k\in\tilde{\mathbf{g}}}b_k(Z(\sigma_k))\,\e^{e_k\,g(\sigma_k)}\,,
\ee
where $(i\,j):=\epsilon_{\mathbf{ba}}\sigma_i^{\mathbf{a}}\sigma_j^{\mathbf{b}}$, $\D\sigma:=(\sigma\,\d\sigma)$ and 
\be\label{hdegsplit}
g(\sigma):=\frac{1}{2\pi\im}\int_{\P^1}\frac{\D\sigma'}{(\sigma\,\sigma')}\,\frac{(\iota\,\sigma)}{(\iota\,\sigma')}\,\sa(Z(\sigma))\,,
\ee
for $\sa(Z)$ the partial connection of the SDD on twistor space. While this formula is mathematically well-defined (in the sense that all integrals make projective sense and the formula has the correct little group weights in the external momenta), we have not yet been able to perform any of the moduli integrals explicitly for $k>0$. Furthermore, unlike our MHV amplitude, there is no first-principles derivation of the expression \eqref{NkMHV}, although it can be obtained from a worldsheet correlator in a background-coupled twistor string theory. It would be very interesting to further simplify the formula \eqref{NkMHV}, as well as to provide tests or a proof of its veracity.

Finally, in~\cite{Adamo:2023fbj} we also gave quasi-momentum eigenstates on the curved \emph{spacetime} which is the classical double copy of the SDD: namely, the self-dual Taub-NUT metric. It is possible to compute the MHV graviton scattering amplitude on a SD Taub-NUT spacetime using a gravitational analogy of the methods utilized here, although there are significant new complexities which arise due to the non-linearity of the background metric. We will report on this gravitational calculation in upcoming work.

	\acknowledgments
	We thank Simon Heuveline and David Skinner for helpful conversations. TA is supported by a Royal Society University Research Fellowship, the Leverhulme Trust grant RPG-2020-386, the Simons Collaboration on Celestial Holography MPS-CH-00001550-11 and the STFC consolidated grant ST/X000494/1. GB is supported by a joint Clarendon Fund and Merton College Mathematics Scholarship. LJM would like to thank the Institute des Haut \'Etudes Sci\'entifique, Bures Sur Yvette, and the Laboratoire Physique at the ENS, Paris for hospitality while this was being written up, and the Simons Collaboration on Celestial Holography MPS-CH-00001550-11 and STFC for financial support from grant number ST/T000864/1. AS is supported by a Black Hole Initiative fellowship, funded by the Gordon and Betty Moore Foundation and the John Templeton Foundation. 
	
	
	\appendix
	\section{Integration of the ``single-minus'' MHV amplitude}\label{app:integration}
    In this Appendix, we show how to derive the integrated amplitudes \eqref{eq:integrated_2point}, \eqref{eq:integrated_3point}, \eqref{eq:integrated_npoint} from the general integral formula \eqref{eq:minimal_amplitude}. Let us focus on the $n$-point amplitude and suppose $e_1<0<e_2,\ldots, e_n$. After rescaling $r\mapsto r(1+\zeta \bar\zeta)$, the amplitude is
    \be
        A_n=4\pi\im g^{n-2}\delta(\omega)\delta(e)\frac{\la r\,s\ra^4}{\la 1\,2\ra\ldots\la n\,1\ra}\mathcal{I}_n\,,
    \ee
    where
    \be
    \mathcal{I}_n\coloneqq \int_0^\infty\d r\,r^2\int_\C\d \zeta\,\d\bar\zeta(1+\zeta\bar\zeta)\exp\left(\im k^3r(1-\zeta\bar\zeta+\bar \zeta w+\zeta\bar w\right)\prod_{j=2}^n[r(\zeta-z_1)(\bar \zeta z_j+1)]^{2e_j}\,.
    \ee
    Here the total 3-momentum $\vec k=\vec k_1+\cdots +\vec k_n$ is expressed in terms of its projection $k^3$ along the $z$ axis and
    \be
        w\coloneqq \frac{k^1+\im k^2}{k^3}\,,\qquad \bar w\coloneqq\frac{k^1-\im k^2}{k^3}\,.\label{eq:w_barw}
    \ee
    Let us now regularize the integral via an $\im\eps$ prescription and also introduce $n-1$ parameters $\rho_i$ in order to express the integral as a Gaussian integral
    \be
    \mathcal{I}_n=\lim_{\rho_j,\eps\to0}\prod_{j=2}^n(-\im \p_{\rho_j})^{2e_j}\int_0^\infty\d r\,r^2\int_\C\d \zeta\,\d\bar\zeta(1+\zeta\bar\zeta)\e^{\im r(A\zeta\bar\zeta+B\zeta+C\bar\zeta+D)-\eps r}\,,
    \ee
    where
    \be
    \begin{aligned}
        A&=-k^3+\sum_{j=1}^n\rho_jz_j\,, \qquad B=k^3\bar w+\sum_{j=2}^n\rho_j\,,\\ C&=k^3 w-\sum_{j=2}^n\rho_jz_1z_j\,,\qquad D=k^3-\sum_{j=2}^n\rho_jz_1
    \end{aligned}
    \ee
    The integration is now straightforward and leads to
    \be
    \begin{aligned}
        \mathcal{I}_n&=\pi\lim_{\rho_j\to0}\prod_{j=2}^n(-\im\p_{\rho_j})^{2e_j}\frac{A+D}{(AD-BC)^2}\\&=\pi\lim_{\rho_j\to0}\prod_{j=2}^n(-\im\p_{\rho_j})^{2e_j}\frac{\displaystyle\sum_{k=2}^n\rho_kz_{k1}}{\left(\vec k^2+\displaystyle\sum_{i=2}^n\alpha_{i1;n}\rho_i\right)^2}\,.
    \end{aligned}
    \ee
    We took the limit $\eps\to0$ by discarding any distributional term, that is neglecting any contribution to forward scattering. In going to the second line, we expressed the total momentum $\vec k^2=(k^3)^2(1+w\bar w)$ and noticed the identity
    \begin{equation}
        \alpha_{ij;n}=k^3(w-z_i-z_j-\bar wz_iz_j)\,,
    \end{equation}
    which follows directly from \eqref{eq:spinor_momenta} and \eqref{eq:w_barw}. The derivative can be extracted via the Taylor series of the right-hand side and leads to
    \be
    \mathcal{I}_n=\frac{\pi(-)^{2e_2+\ldots 2e_n}(2e_2+\ldots+2e_n)!}{|\vec k_1+\ldots\vec k_n|^{2+4e_2+\ldots +4e_n}}\prod_{j=2}^n(\im\alpha_{j1;n})^{2e_j}\sum_{k=2}^n\frac{2e_kz_{k1}}{\alpha_{k1;n}}\,.
    \ee
    On the support of charge conservation, we finally reconstruct \eqref{eq:integrated_2point}, \eqref{eq:integrated_3point} and \eqref{eq:integrated_npoint}.
	
	
	\bibliographystyle{JHEP}
	\bibliography{sdc}
\end{document}